\documentclass{Interspeech}
\usepackage{booktabs}  
\usepackage{tabularx}  
\usepackage{array}  
\usepackage{makecell} 
\usepackage{multirow} 

\interspeechcameraready

\title{Cross-attention and Self-attention for Audio-visual Speaker Diarization in MISP-Meeting Challenge}

\author[affiliation={1}]{Zhaoyang}{Li}
\author[affiliation={1}]{Haodong}{Zhou}
\author[affiliation={1}]{Longjie}{Luo}
\author[affiliation={2}]{XiaoXiao}{Li}
\author[affiliation={1}]{Yongxin}{Chen}
\author[affiliation={1}]{Lin}{Li*}
\author[affiliation={3}]{Qingyang}{Hong*}

\affiliation{School of Electronic Science and Engineering}{Xiamen University}{China}
\affiliation{School of Electronic Information}{Beijing Jiaotong University}{China}
\affiliation{School of Informatics}{Xiamen University}{China}
\email{23120231150269@stu.xmu.edu.cn, \{lilin,qyhong\}@xmu.edu.cn}

\usepackage{comment}

\begin{document}

\maketitle

\renewcommand{\thefootnote}{\fnsymbol{footnote}}
\footnotetext[1]{Corresponding author.}
\renewcommand{\thefootnote}{\arabic{footnote}}
\begin{abstract}
This paper presents the system developed for Task 1 of the Multi-modal Information-based Speech Processing (MISP) 2025 Challenge.
We introduce CASA-Net, an embedding fusion method designed for end-to-end audio-visual speaker diarization (AVSD) systems. CASA-Net incorporates a cross-attention (CA) module to effectively capture cross-modal interactions in audio-visual signals and employs a self-attention (SA) module to learn contextual relationships among audio-visual frames. To further enhance performance, we adopt a training strategy that integrates pseudo-label refinement and retraining, improving the accuracy of timestamp predictions. Additionally, median filtering and overlap averaging are applied as post-processing techniques to eliminate outliers and smooth prediction labels. Our system achieved a diarization error rate (DER) of 8.18\% on the evaluation set, representing a relative improvement of 47.3\% over the baseline DER of 15.52\%.
\end{abstract}

\section{Introduction}

Audio-visual speaker diarization (AVSD) is a technology that integrates both audio and visual signals to determine  ``who spoke when" \cite{park2022review} in multi-speaker conversational scenarios. It has been widely applied in areas such as remote video conferencing summaries \cite{ryant2020third} and audiovisual speech transcription \cite{fujita2019end}.

AVSD has evolved from traditional audio-based speaker diarization (ASD) \cite{medennikov2020target}. In complex environments, audio signals are often affected by external noise, reverberation, and speaker overlap, hindering the effectiveness of single-modal ASD systems. Given that visual cues enhance human speech perception, researchers have explored the integration of video information to improve speaker diarization beyond audio-only approaches \cite{yehia1998quantitative, mcgurk1976hearing}.
To address this, the MISP 2025 Challenge introduced Task 1, which specifically focuses on AVSD by assigning speech timestamps based on speaker identity. This challenge aims to improve meeting transcription by leveraging multimodal information, including video data.

The research presented in \cite{he2022end} indicates that Visual Voice Activity Detection (V-VAD) outperforms traditional audio-only speaker diarization in handling overlapped speech segments in the MISP corpus. V-VAD is an advanced approach that leverages visual cues to detect speech onset, offset, and presence by analyzing facial expressions, lip movements, and other visual indicators extracted from video frames \cite{ma2021lip}.
Additionally, incorporating temporal context information from video data \cite{wang2022efficient, wei2022spatial} enhances the learning of robust visual representations. However, the effectiveness of V-VAD may degrade in scenarios where visual features are obstructed by occlusions or when speakers are off-screen. Moreover, the accuracy of lip movement detection can be affected by head movements or environmental conditions, whereas audio signals remain resilient under such conditions. Thus, developing techniques that effectively capture the relationship between audio and lip movements, along with contextual information across audio-visual frames, is crucial for improving the performance of audio-visual speaker diarization.

We propose an audio-visual speaker diarization system for Task 1 of the MISP 2025 Challenge, focusing on effectively integrating audio-visual modalities and improving the accuracy of temporal label predictions. Our approach employs a Cross-Attention (CA) module to dynamically align feature sequences between audio and visual modalities, addressing temporal misalignment between video and audio data. This is followed by a Self-Attention (SA) module to learn contextual relationships between temporal frames. To further enhance performance, we utilize visual features to generate pseudo-labels for speaker timestamps and retrain the network, iteratively improving the system's pseudo-label correction capability. Additionally, Mixup data augmentation is applied to both lip images and speaker embeddings to prevent overfitting. Finally, we apply median filtering and overlap averaging as post-processing techniques to eliminate anomalous predictions and smooth the predicted labels.

Using CASA-Net \cite{zhou2023casa}, we achieved a diarization error rate (DER) of 7.35\% on the development set and 8.18\% on the evaluation set, outperforming the baseline by 47.3\% in Task 1 of the MISP 2025 Challenge.

The remainder of this paper is organized as follows: Section 2 reviews related work, including baseline models, encoders, extractors, and decoders. Section 3 describes the proposed network framework. In Section 4, we present the experimental setup, followed by result analysis in Section 5. Finally, Section 6 concludes the research.

\section{Baseline works}
\begin{figure}[t]
  \centering
  \includegraphics[width=\linewidth,height=6cm]{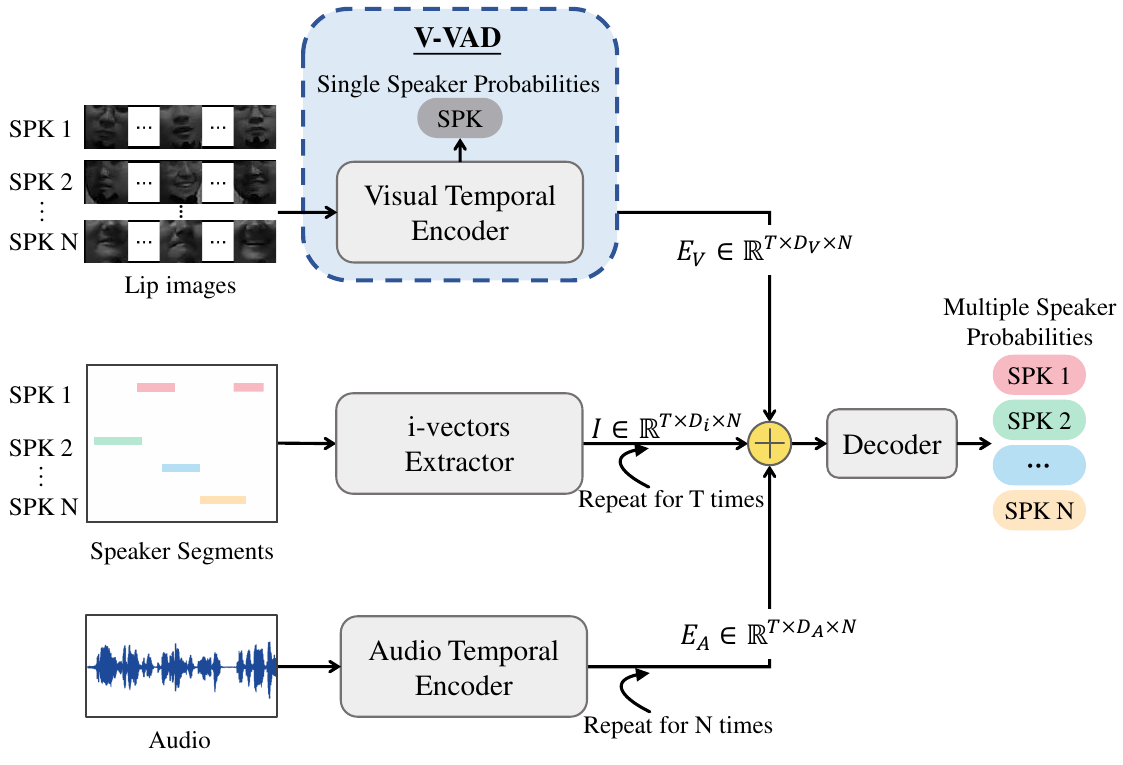}
  \caption{The baseline of MISP 2025 Challenge}
  \label{fig:misp_baseline}
\end{figure}

The MISP 2025 Challenge employs a baseline model \cite{he2022end} that integrates several key components: visual and audio temporal encoders, an i-vector extractor, and a decoder. The overall architecture is depicted in Figure \ref{fig:misp_baseline}. The following subsections provide detailed descriptions of each module.

\subsection{Visual temporal encoder}
The visual temporal encoder functions as a V-VAD module. The input video is divided into multiple segments, each containing lip images. These images are processed by the visual temporal encoder, formulated as \( X_V \in \mathbb{R}^{T \times W \times H \times N} \), where \( T \) denotes the number of frames, and \( W\), \(H \) represent the image width and height, respectively, while \( N \) is the number of speakers. The encoder extracts visual embeddings \( E_V \in \mathbb{R}^{T \times D_V \times N} \), where \( D_V \) is the embedding dimension.

\subsection{Audio temporal encoder}
The audio temporal encoder processes FBank features through a four-layer convolutional neural network (CNN), where each layer comprises 2D convolution, batch normalization (BatchNorm), and a ReLU activation function. The CNN output is passed through a fully connected layer, yielding an intermediate representation \( \hat{E}_A \in \mathbb{R}^{T \times D_A} \). To align with the visual embeddings \( E_V \), this representation is replicated \( N \) times, resulting in the final audio embeddings \( E_A \in \mathbb{R}^{T \times D_A \times N} \).

\subsection{I-vector extractor}
To mitigate alignment challenges arising from occlusions or imprecise lip detection, an i-vector approach is employed. A 100-dimensional i-vector extractor, pre-trained on the CN-Celeb dataset \cite{fan2020cn}, generates speaker embeddings \( \hat{I} \in \mathbb{R}^{D_I \times N} \). These embeddings are then replicated \( T \) times to form \( I \in \mathbb{R}^{T \times D_I \times N} \), ensuring temporal alignment with both visual and audio embeddings.

\subsection{Decoder}
The decoder fuses the visual, audio, and i-vector embeddings into a unified feature representation \( E_{\text{all}} \in \mathbb{R}^{T \times (D_A + D_V + D_I) \times N} \). The decoder processes this representation to generate the final diarization output \( S \in \mathbb{R}^{T \times N} \). A softmax function is applied to compute class probabilities, followed by a loss function that measures the discrepancy between predicted outputs and ground truth labels, facilitating model optimization.

\begin{figure*}[t]
  \centering
  \includegraphics[width=\linewidth,height=8.9cm]{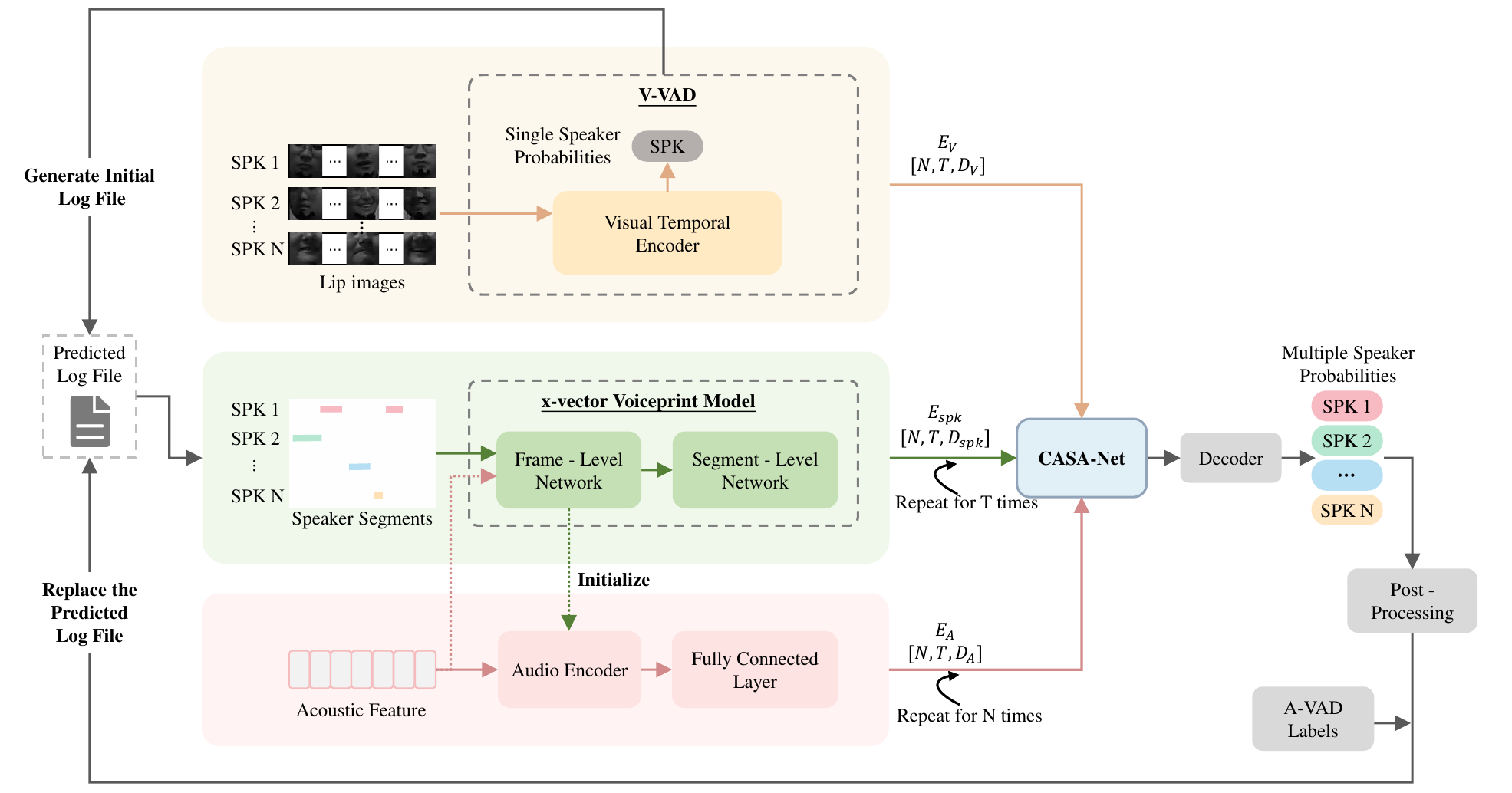}
  \caption{Framework of the audio-visual speaker diarization system based on CASA-Net.}
  \label{fig:overpipline}
\end{figure*}
\section{System description}
The overall audio-visual speaker diarization system based on CASA-Net is shown in Figure \ref{fig:overpipline}. Compared to the baseline, the main improvements include modifications to the visual encoder, the introduction of the CASA-Net architecture, and the incorporation of an x-vector-based speaker embedding model as both the audio encoder and speaker embedding extractor. Additionally, the training strategy has been optimized, and effective techniques such as Mixup data augmentation, median filtering, and overlapping averaging have been applied as post-processing methods. The following sections provide a detailed explanation of these enhancements.

\subsection{Cross-Attention (CA) and Self-Attention (SA)}
\begin{figure}[t]
  \centering
  \includegraphics[width=\linewidth,height=8cm]{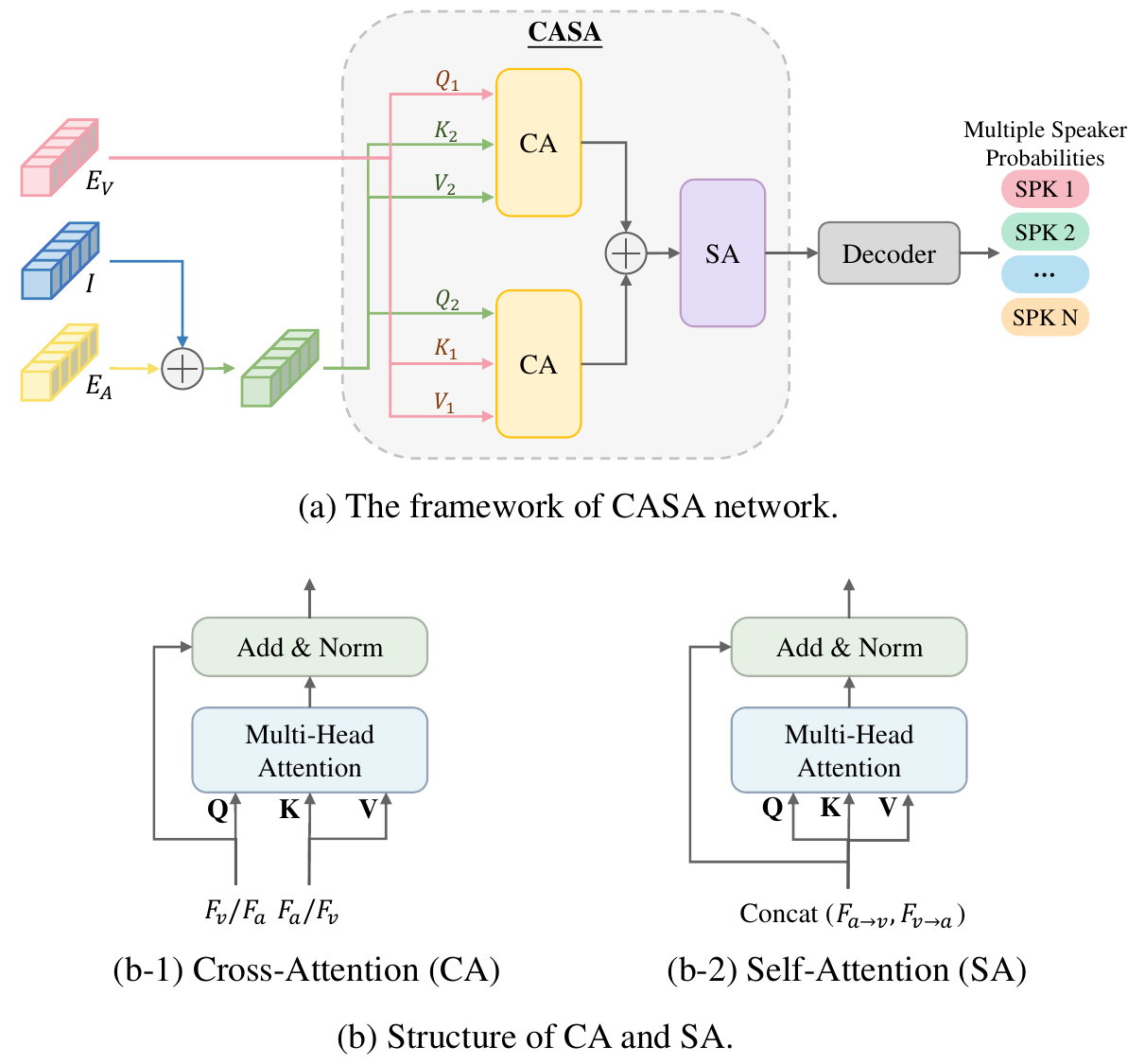}
  \caption{The framework of CASA network and the structure of CA and SA}
  \label{fig:ca_sa_s}
  \vspace{-2em} 
\end{figure}

The architecture of the proposed CASA-Net is illustrated in Figure \ref{fig:ca_sa_s}(a). The system takes lip images, i-vectors, and audio signals as inputs, which are processed via the visual temporal encoder, i-vector extractor, and audio temporal encoder to obtain their respective feature embeddings. Given that i-vectors and audio features both pertain to the audio modality, we combine them to enhance complementary information. Instead of the concatenation approach adopted in \cite{he2022end}, we introduce the CASA network, which integrates a cross-attention (CA) module and a self-attention (SA) module. These modules facilitate the learning of intrinsic cross-modal relationships among the extracted embeddings.

Although achieving perfect temporal alignment of audio-visual data is challenging, we aim to capture momentary interactions between audio and visual signals at the frame level via cross-attention. Specifically, the audio and visual features serve as the query (\(Q\)), key (\(K\)), and value (\(V\)) in our audio-visual cross-attention framework.

Since both \(\mathbf{E}_{A}\) and \(\mathbf{I}\) are derived from audio information, we concatenate them into \(\mathbf{F}_{a} \in \mathbb{R}^{T \times (D_A + D_I) \times N}\), while the visual embeddings \(\mathbf{E}_V\) form \(\mathbf{F}_{v} \in \mathbb{R}^{T \times D_V \times N}\). A multi-head attention mechanism is employed to enhance feature representation learning by attending to multiple aspects of the input data. As shown in Eq.~(\ref{CAav}) and (\ref{CAva}), when generating \(F_{a \rightarrow v} \in \mathbb{R}^{T \times D_V \times N}\), \(\mathbf{F}_{v}\) acts as the query and \(\mathbf{F}_{a}\) as the key/value; conversely, the roles are reversed for \(F_{v \rightarrow a}\). The left side of Figure \ref{fig:ca_sa_s}(b) presents the detailed structure of \(F_{a \rightarrow v}\)\verb|/|\(F_{v \rightarrow a}\).
\begin{align}
F_{a \rightarrow v} &= \operatorname{softmax}\left(\frac{Q_v K_a^T}{\sqrt{d}}\right) V_a
\label{CAav}
\end{align}

\begin{align}
F_{v \rightarrow a} &= \operatorname{softmax}\left(\frac{Q_a K_v^T}{\sqrt{d}}\right) V_v
\label{CAva}
\end{align}
The self-attention (SA) module is designed to capture temporal dependencies within audio-visual frames and strengthen the correlations among them. As illustrated in Figure \ref{fig:ca_sa_s}, the SA module receives its query, key, and value inputs by concatenating \(F_{a \rightarrow v}\) and \(F_{v \rightarrow a}\), thereby enabling context-aware feature learning.
\subsection{X-vector extractor and pseudo-label refinement}
To mitigate ambiguities in timestamp assignment for overlapping speech segments, we introduce an x-vector-based ECAPA-TDNN \cite{desplanques2020ecapa} speaker embedding pre-trained model as both the audio encoder and speaker embedding extractor. Specifically, the audio encoder shares the same architecture as the frame-level network, while the x-vector extracted from the segment-level network replaces the i-vector in the baseline system as the speaker embedding feature.

To address the mismatch of speaker embedding features between the training and testing phases, we employ a model training strategy based on pseudo-label correction and retraining. The process begins with extracting speaker embedding features from an initial log file generated using V-VAD for the first-round training. The system then computes the loss based on the predicted results and ground-truth labels. Through post-processing, timestamp pseudo-labels are generated and converted into a new log file, replacing the initial log file. This process is iteratively refined over multiple training rounds.
\subsection{Mixup Data Augmentation and Post-processing}
Mixup \cite{zhang2017mixup} is a data augmentation technique grounded in Vicinal Risk Minimization (VRM) \cite{el2024augment}. It synthesizes new training instances by linearly interpolating multiple samples along with their corresponding labels. To mitigate overfitting, we apply Mixup to linearly blend lip images, speaker embeddings, and corresponding labels from different speakers in the training dataset.

To refine the predicted sequence, we employ median filtering \cite{huang1981two} to preserve edge integrity while removing isolated outliers. Additionally, to mitigate key information loss and abrupt changes in predicted values due to signal truncation at window boundaries, we adopt a sliding window overlapping averaging technique \cite{dehghani2019quantitative}. This approach helps maintain essential signal characteristics across overlapping windows, thereby improving prediction stability.

\section{Datasets and experimental setup}
The overall audio-visual speaker diarization system based on CASA-Net is shown in Figure \ref{fig:overpipline}. The following section details the experimental process.
\subsection{Data preparation}

Our primary dataset is the MISP 2025 Challenge corpus, which comprises a 119-hour training set (each meeting lasting 2 hours), a 3-hour development set (each meeting lasting 20 minutes), and an additional 3-hour evaluation set for final assessment. To further improve model robustness, we incorporate the MISP 2022 dataset as supplementary training data.

The dataset undergoes several preprocessing steps to enhance audio and visual quality. First, we employ NARA-WPE \cite{drude2018nara} for dereverberation of multi-channel audio. To leverage spatial information from multiple channels, the Adaptive Beamforming algorithm from the Kaldi toolkit \cite{povey2011kaldi} is applied to merge multi-channel far-field audio into a single-channel signal. The augmented audio dataset consists of 8-channel WPE-dereverberated audio, a beamformed single-channel audio, and 8-channel raw audio, resulting in a total of 17 audio channels.

For the image data, negative sampling is applied to the training set by incorporating non-speaking lip images and out-of-session speaker embeddings, ensuring that each session fragment meets the maximum number of speakers. Subsequently, Mixup augmentation is applied to sequential lip images, speaker embeddings, and corresponding labels across all speakers within a session.

\subsection{Implementation details}
\begin{itemize}
\item \textbf{Training}: During training, audio and video signals are segmented into 8-second blocks with a 4-second overlap. For the audio input, 40-dimensional filter banks (FBanks) are extracted as acoustic features, with a frame length of 25 ms and a frame shift of 10 ms, yielding 800 audio frames per block. In the video modality, lip region images (\(96 \times 96\) resolution) are extracted at 25 FPS, resulting in 200 visual frames per block.

 Initially, the V-VAD model is trained using binary cross-entropy (BCE) loss and the Adam optimizer with a learning rate of \(1 \times 10^{-4}\). After training, the model is frozen, and frame-wise video features are extracted to serve as input for the video modality. Next, audio inputs are processed and input into the CASA network and decoder. The BCE loss function and Adam optimizer are employed, maintaining a learning rate of \(1 \times 10^{-4}\) until convergence.

\item \textbf{Evaluation}: During evaluation, all development set recordings were segmented into 8-second blocks with a 4-second stride. Final scores were computed by averaging predictions from overlapping regions of adjacent blocks. Additionally, the competition provided access to the Oracle VAD, which participants were permitted to utilize.
\subsection{Metrics}
We measure the performance of our model by DER, which is represented with Eq. (\ref{DER}).
\begin{align}
DER=\frac{FA+MISS+SpkErr}{TOTAL}
  \label{DER}
\end{align}
where \(FA\) is the speech durations of false alarm, \(MISS\) is the speech durations of missed detection, \(SpkErr\) is the speech durations of speaker error, and \(TOTAL\) is the sum of durations of all reference speakers’ utterances.
\end{itemize}

\section{Experimental results}

To quantitatively evaluate the contribution of each proposed component, we conduct a comprehensive ablation study under the DIHARD-III evaluation protocol. As systematically summarized in Table \ref{tab:method_comparison}, the experimental results reveal critical insights about the structure design of our proposed system. Our final integrated system achieves a DER of 8.18\%, which not only matches the state-of-the-art performance on the leaderboard but also demonstrates a remarkable 47.3\% relative improvement (7.34\% absolute reduction) over the official baseline DER of 15.52\%. The observed progressive performance degradation following component removal further validates our hierarchical design philosophy.

Removing the post-processing module, including median filtering and overlapping averaging, results in a 0.85\% performance drop, indicating its effectiveness in handling anomalies and smoothing boundary predictions for greater stability. Excluding Mixup leads to a 0.68\% decline, demonstrating its role in preventing overfitting and improving generalization. Eliminating pseudo-label refinement causes a 0.7\% degradation, highlighting its importance in correcting timestamp labels during testing and addressing speaker embedding discrepancies caused by varying timestamp accuracy between training and testing. Additionally, removing ECAPA-TDNN reduces performance by 0.59\%, confirming that x-vector provides superior speaker identity representation compared to i-vector, further enhancing system performance.

Finally, the proposed CASA-Net feature fusion network is a key component in boosting system performance. It effectively mitigates temporal misalignment between audio and video streams while capturing long-range dependencies, leading to more accurate speaker label predictions.

\section{Conclusion}
In this paper, we propose an audio-visual speaker diarization system  for Task 1 of the MISP 2025 Challenge. Our system employs a CASA-Net based architecture to effectively address the temporal misalignment between audio and video streams. By integrating cross-attention and self-attention modules, we achieve seamless audio-visual feature fusion while capturing global temporal dependencies. Additionally, we leverage an x-vector speaker embedding pre-trained model  as a feature extractor to enhance speaker identity representation. System performance is further optimized through optimized training strategies, Mixup data augmentation, and post-processing techniques. Our final system achieved a DER of 8.18\% on the evaluation set, representing a relative improvement of 47.3\% compared to the baseline DER of 15.52\%.
\begin{table}[t!]
    \centering
    \caption{DER(\%) on the Development and Evaluation Sets: Performance comparison of different methods}
    \label{tab:method_comparison}
    
    \begin{tabular}{l c c}
        \toprule
        \textbf{Method} & \textbf{Dev Set} & \textbf{Eval Set} \\
        \midrule
        Official Baseline & - & 15.52\\
        \textbf{Our System} & 7.35 & \textbf{8.18} \\
        \hspace{5pt} - Post-processing & 8.45 & 9.03 \\
        \hspace{5pt} - Mixup & 9.04 & 9.71 \\
        \hspace{5pt} - Pseudo-label refinement & 9.92 & 10.41 \\
        \hspace{5pt} - ECAPA-TDNN & 10.64 & 11.00 \\
        \hspace{5pt} - CASA-Net & 15.98 & 17.04 \\
        \bottomrule
    \end{tabular}
    \vspace{-0.4cm}
\end{table}
\section{Acknowledgements}
This work was supported in part by the National Natural Science Foundation of China under Grants 62371407 and 62276220, and the Innovation of Policing Science and Technology, Fujian province (Grant number: 2024Y0068)

\bibliographystyle{IEEEtran}
\bibliography{main}

\end{document}